\documentstyle[prl,aps,epsf]{revtex}

\begin{document}
\title{Cascades and statistical equilibrium in shell models of 
turbulence}
\author{P. D. Ditlevsen and I. A. Mogensen\\
The Niels Bohr Institute, Department for Geophysics,\\
University of Copenhagen,\\
Haraldsgade 6, DK-2200 Copenhagen N, Denmark.}
\date{July 9, 1995}
\maketitle

\begin{abstract}
We study the GOY shell model simulating the cascade processes of
turbulent flow. The model has two inviscid invariants governing the
dynamical behavior. Depending on the choice of interaction
coefficients, or coupling parameters, the two invariants are either
both positive definite, analogous to energy and enstrophy of 2D flow,
or only one is positive definite and the other not, analogous to energy
and helicity of 3D flow.  In the 2D like model the dynamics depend on
the spectral ratio of enstrophy to energy. That ratio depends on
wave-number as $k^{\alpha}$. The enstrophy transfer through the
inertial sub-range can be described as a forward cascade for $\alpha <
2$ and a diffusion in a statistical equilibrium for $\alpha > 2$. The
$\alpha =2$ case, corresponding to 2D turbulence, is a borderline
between the two descriptions. The difference can be understood in
terms of the ratio of typical timescales in the inertial sub-range
and in the viscous sub-range. The multi fractality of the enstrophy
dissipation also depends on the parameter $\alpha$, and seems to be
 related to
the ratio of typical timescales of the different shell velocities.
\end{abstract} 

\section{Introduction}
The standard Kolmogorov $k^{-5/3}$ scaling law for energy cascading in
3D turbulence and the corresponding $k^{-3}$ scaling law for enstrophy
cascade in 2D turbulence is still debated.  Direct numerical
calculations of the full Navier-Stokes equation is by and large still
impossible for high Reynolds number $( > 100-200)$ flows.  However, the
cascading mechanisms and its multi fractal nature can be analyzed in
reduced wave-number models for very high Reynold numbers with high
accuracy.  In this paper we investigate the GOY shell
model \cite{Gledzer,GOY} where the spectral velocity or vorticity is
represented by one complex variable for each shell evenly spaced in
$log(k)$ in spectral space.  For this type of model the Kolmogorov
scaling arguments can be applied as for real flow regardless of how
realistically they mimic the dynamics of the Navier-Stokes equation.
The scaling behavior of the fields depends on the inviscid invariants
of the model.  In the simple model we are able to control which
symmetries and conserved integrals of the dynamics that are present in
the inviscid and force-free limit. In the models we interpret as
simulating 3D turbulence there are 2 inviscid invariants, similar to
energy and helicity \cite{benzi}, of which the first is positive definite and the
second is not. For the models we interpret as 2D turbulence the 2
inviscid invariants, similar to energy and enstrophy, are both positive
definite. We will mainly be concerned with an investigation of the 2D
like models.  The specific parameter choice previously assigned to
simulating 2D turbulence are such that the GOY model does not show
enstrophy cascading but rather a statistical equilibrium where the
enstrophy is transported through the inertial sub-range by
diffusion \cite{aurell}. We show that this is a borderline case for
which, on one side, the model behaves as a cascade model and, on the
other side, it behaves as a statistical equilibrium model, where the
enstrophy spectrum is characterized by a simple equipartitioning among
the degrees of freedom of the model. The difference in behavior is
connected with the different typical timescales of the shell velocities
as function of shell number. This probabily also influences the (non-universal)
multi fractal behavior of the shell velocities. If timescales in the
viscous sub-range are not smaller than in the beginning
of the inertial sub-range,the low wave-number end, the model does not have a
multi fractal spectrum.

\section{The GOY model}  
The GOY model is a simplified analogy to the spectral Navier-Stokes
equation for turbulence. The spectral domain is represented as shells,
each of which is defined by a wavenumber $k_n = k_0 \lambda^n$, where
$\lambda$ is a scaling parameter defining the shell spacing; in our
calculations we use the standard value $\lambda=2$.  The reduced phase
space enables us to cover a large range of wavenumbers, corresponding
to large Reynolds numbers.  We have $2 N$ degrees of freedom, where $N$
is the number of shells, namely the generalized complex shell
velocities or vorticities, $u_n$ for $n=1,N$. The dynamical equation
for the shell velocities is,

\begin{equation}
\dot{u}_n=i k_n (a 
u^*_{n+2}u^*_{n+1}+\frac{b}{\lambda}u^*_{n+1}u^*_{n-1}
+\frac{c}{\lambda^2}u^*_{n-1}u^*_{n-2}) -\nu k_n^{p_1} u_n - \nu' 
k_n^{-p_2} u_n+ f_n,
\label{dyn}
\end{equation}
where the first term represents the non-linear wave interaction or
advection, the second term is the dissipation, the third term is a
drag term, specific to the 2D case, and the fourth term is the forcing.  Throughout this paper
we use $p_1=p_2=2$.  We will for convenience set $a=k_0=1$, which can
be done in (\ref{dyn}) by a rescaling of time and the units in $k$ space.  
A real form of the GOY
model, as originally proposed by Gledzer \cite{Gledzer}, can be
obtained trivially by having purely imaginary velocities and forcing.
The GOY model in its original real form contains no information about phases between waves, thus
there cannot be assigned a flow field in real space to the spectral
field.  The complex form of the GOY model and extensions in which there
are more shell variables in each shell introduce some degrees of
freedom, which could be thought of as representing the phases among
waves.  However, it seems as if these models do not behave differently
from the real form of the model in regard to the conclusions in the
following \cite{euro,aurell}. The key issue for the behavior of the
model is the symmetries and conservation laws obeyed by the model.

\subsection{Conservation laws}

The GOY model has two conserved integrals, in the case of no forcing
and no dissipation $(\nu = f = 0)$.
We denote the two conserved integrals by

\begin{equation}
E^{1,2}=\sum_{n=1}^N E^{1,2}_n=\frac{1}{2}\sum_{n=1}^N 
k_n^{\alpha_{1,2}}|u_n|^2=
\frac{1}{2}\sum_{n=1}^N \lambda^{n\alpha_{1,2}}|u_n|^2
\end{equation}

By setting $\dot{E}^{1,2}=0$ and using $\dot{u}_n$ from (\ref{dyn}) 
we get

\[ \left\{ \begin{array}{ll}
1 + b z_1 + c z_1^2 = 0 \\ 
1 + b z_2 + c z_2^2 = 0,
\end{array} \right.  \]
\begin{equation}\label{bc}\end{equation}
where the roots $z_{1,2}=\lambda^{\alpha_{1,2}}$ are the generators
of the conserved integrals. In the case of negative values of $z$ we
can use the complex formulation, $\alpha=(log|z|+i\pi)/log\lambda$.
The parameters $(b, c)$ are determined from (\ref{bc}) as

\[ \left\{ \begin{array}{ll}
b=-(z_1+z_2)/z_1z_2  \nonumber \\
c=1/z_1z_2.
\end{array} \right.  \]
\begin{equation}\label{2}\end{equation}
In the $(b,c)$ parameter plane the curve $c=b^2/4$ represents models
with only one conserved integral, see figure 1. Above the parabula the
generators are complex conjugates, and below they are real and
different. Any conserved integral represented by  a real nonzero
generator $z$ defines a line in the $(b, c)$ parameter plane, which is
tangent to the parabula in the point $(b,c)=(-2/z,1/z^2)$. The rest of
our analysis we will focus on the line defined by $z_1=1$. The
conserved integral,

\begin{equation}
E^1=\frac{1}{2} \sum_{n=1}^N |u_n|^2,
\end{equation}
is the usual definition of the energy for the GOY model \cite{benzi}.
The parameters are then determined by $1+b+c=0$, which with the
definitions $b=-\epsilon$ and $c=-(1-\epsilon )$ agree with the notation
of ref.  \cite{biferale}.  The generator of the other conserved
integral is from (\ref{2}) given as,

\begin{equation}
z_2=\frac{1}{\epsilon-1}.
\label{z2}
\end{equation}

For $\epsilon < 1$ the second conserved integral is not positive
definite and is of the form,

\begin{equation}
E^2=H=\frac{1}{2} \sum_{n=1}^N (-1)^n |z_2|^n |u_n|^2,
\label{helicity}
\end{equation}
which can be interpreted as a generalized helicity. For $\epsilon=1/2$,
$z_2 =-2=-\lambda$ the model is the usual 3D shell model and  H is the
helicity as defined in ref. \cite{benzi}.  By choosing $\lambda$ such
that $\lambda = 1/(1-\epsilon)$ we get $E^2= \sum (- 1)^n\lambda^n
|u_n|^2$. This form was argued in ref. \cite{benzi} to be the proper
form for the helicity. In this paper we will alternatively use the
definition (\ref{helicity}) for the helicity.

For $\epsilon >
1$ the second conserved integral is positive definite and of the form,

\begin{equation}
E^2=Z=\frac{1}{2}\sum_{n=1}^N z_2^n |u_n|^2,
\end{equation}
which can be interpreted as a generalized enstrophy. For
$\epsilon=5/4$, $z_2=4 =\lambda^2$ the model is the usual 2D shell
model and  Z is the enstrophy  as defined in ref. \cite{aurell}.  The
sign of $c$, which is the interaction coefficient for the smaller
wavenumbers, changes when going from the 3D - to the 2D case. This
could be related to the different role of backward cascading in the two
cases. To see this, consider the non-linear transfer of $E^i$ through
the triade interaction between shells, $n-1,n,n+1$. This is simply
given by,

\begin{eqnarray}
\dot{E}^i_{n-1}&=&k^{\alpha_i}_{n-1}\Delta_n \nonumber \\
\dot{E}^i_{n}&=&bz_ik^{\alpha_i}_{n-1}\Delta_n\nonumber  \\
\dot{E}^i_{n+1}&=&cz_i^2k^{\alpha_i}_{n-1}\Delta_n,
\end{eqnarray}
with

\begin{equation}
\Delta_n=k_{n-1}Im(u_{n-1}u_nu_{n+1}).
\end{equation}

The detailed conservation of $E^i$ in the triade interaction is
reflected in the identity, $1+bz_i+cz_i^2=0$. Using (\ref{2}) and 
(\ref{z2}) we have for the exchange of energy, $E^1$, with 
$\alpha_1=0$;

\begin{eqnarray}
\dot{E}^1_{n-1}&=&\Delta_n\nonumber   \\
\dot{E}^1_{n}&=&-\epsilon\Delta_n\nonumber   \\
\dot{E}^1_{n+1}&=&(\epsilon-1)\Delta_n 
\end{eqnarray}

and for helicity/enstrophy, $E^2$, with $\alpha_2=\alpha$;

\begin{eqnarray}
\dot{E}^2_{n-1}&=&k^{\alpha}_{n-1}\Delta_n\nonumber  \\
\dot{E}^2_{n}&=&-(\epsilon/(\epsilon-1))k^{\alpha}_{n-1}\Delta_n \nonumber \\
\dot{E}^2_{n+1}&=&(1/(\epsilon-1))k^{\alpha}_{n-1}\Delta_n. 
\end{eqnarray}

We have $\epsilon<1$ for 3D like models and $\epsilon>1$ for 2D like
models, the two situations are depicted in figure 2, where the
thickness of the arrows symbolize the relative sizes of the exchanges
in the cases of $\epsilon=1/2$ and $\epsilon=5/4$.

\subsection{Scaling and inertial range.}

The inertial sub-range is defined as the range of shells where the
forcing and the dissipation are negligible in comparison with the
non-linear interactions among shells. Since we apply the forcing at the
small shell numbers and the dissipation at the large shell numbers, the
inertial range (of forward cascade) is characterized by the constant
cascade of one of the conserved quantities. The classical Kolmogorov
scaling analysis can then be applied to the inertial range.  There is,
however, in the shell model, long range influences of the dissipation
and forcing into the inertial subrange. This is an artifact of the
modulus 3 symmetry, see equation (\ref{ggg}), and the truncation of the
shell model which is not expected to represent any reality. These
features are treated in great detail in ref. \cite{schorghofer}.
Denoting $\eta_{1,2}$ as the average dissipation of $E^{1,2}$, this is
then also the amount of $E^{1,2}$ cascaded through the inertial range.
The spectrum of $E^{1,2}$ does then, by the Kolmogorov hypothesis, only
depend on $k$ and $\eta_{1,2}$.
From dimensional analysis we have,
$[ku]=s^{-1}$, $[\eta_{1,2}]=[E^{1,2}]s^{-1}$, $[E^{1,2}]=
[k^{\alpha_{1,2}}u^2]=[k]^{\alpha -2}s^{-2}$, and we get,

\begin{equation}
E^{1,2} \sim \eta_{1,2}^{2/3}k^{(\alpha_{1,2} -2)/3}.
\label{k41}
\end{equation}

For the generalized velocity, $u$, we then get the "Kolmogorov-
scaling",

\begin{equation}
|u| \sim \eta_{1,2}^{1/3}k^{-(Re(\alpha_{1,2})+1)/3}.
\end{equation}

The non-linear cascade, or flux, of the conserved quantities defined by $z_{1,2}$
through shell number $n$ can be expressed directly as,

\begin{eqnarray}
\Pi_n^{1,2} = \sum_{m=1}^n\dot{E}^{1,2}(m)&=&\frac{1}{2}\sum_{m=1}^n 
z_{1,2}^m (u^*_m\dot{u}_m+c.c.)\nonumber \\ 
&=&z_{1,2}^n(-\Delta_n /z_{2,1}+
\Delta_{n+1}).
\label{cascade1}
\end{eqnarray}

In the inertial range the cascade is constant,
$\Pi_n^{1,2}=\Pi_{n+1}^{1,2}$, so from (\ref{cascade1}) we get
following ref. \cite{biferale}

\begin{eqnarray}
z_1z_2 \Delta_{n+2}-(z_1+z_2)\Delta_{n+1}+\Delta_{n}=0 \Rightarrow 
\nonumber \\
q_n+z_1z_2/q_{n+1}=z_1+z_2
\label{cascade3}
\end{eqnarray}

where we have defined

\begin{equation}
q_n=\Delta_n/\Delta_{n+1}.
\label{cascade2}
\end{equation}

The inertial range scaling requires $q_n=q_{n+1}=q$ to be
independent of $n$. Solving (\ref{cascade3}) for $q$ and using
(\ref{cascade2}) gives,

\[ q= \left\{ \begin{array}{ll}
z_1 \Rightarrow u_n \sim k_n^{-(\alpha_1+1)/3} & 
\mbox{Kolmogorov for $E^1$} \\ 
z_2 \Rightarrow u_n \sim k_n^{-(\alpha_2+1)/3} & 
\mbox{Kolmogorov for $E^2$}.
\end{array} \right. \label{cascade4} \]

Inserting this into (\ref{cascade1}) gives for the cascade of $E^1$
in the two solutions,

\[ \Pi^1\sim \left\{\begin{array}{ll}
1-z_2/z_1&\mbox{Kolmogorov  for $E^1$}\\ 
0&\mbox{fluxless for $E^1$,} \end{array} \right. \]
\begin{equation} \label{pi1} \end{equation}
and correspondingly for $E^2$,

\[ \Pi^2\sim \left\{\begin{array}{ll}
0&\mbox{fluxless for $E^2$}.\\
1-z_1/z_2&\mbox{Kolmogorov  for $E^2$}.\end{array} \right. \]

These are the two scaling fixed points for the model.  The Kolmogorov
fixed point for the first conserved integral corresponds to the
fluxless fixed point for the other conserved integral and visa versa.
This is of course reflected in the fact that (\ref{cascade3}) is
symmetric in the indices 1 and 2.  That these points in phase space are
fixed points, in the case of no forcing and dissipation, is trivial,
since $\Pi_n=\Pi_{n+1}\Rightarrow \dot{E}_{n+1}=0 \Rightarrow
\dot{u}_{n+1}=0$.  It should be noted that the Kolmogorov fixed point,

\begin{equation}
u \sim k^{-(\alpha +1)/3},
\label{cascade_scaling}
\end{equation}
obtained from this analysis is in agreement with the dimensional
analysis (\ref{k41}).  

The scaling fixed points can be obtained directly from the 
dynamical equation as well. For $u_n \sim k_n^{-\gamma} g(n) =
\lambda^{-n\gamma} g(n)$, where $g(n+3)=g(n)$ is any period 3
function, we get by inserting into (\ref{dyn}) with $a=1$,

\begin{equation}
g(n-1)g(n)g(n+1)\lambda^{n(1-\gamma )+3\gamma }(1+b\lambda^{3\gamma -
1}+c(\lambda^{3\gamma -1})^2)=0
\label{ggg}
\end{equation}
and the generators reemerge,
$z_{1,2}=\lambda^{\alpha_{1,2}}=\lambda^{3\gamma_{1,2} -1}$, giving the
Kolmogorov fixed points for the two conserved integrals, $\gamma_{1,2}=
(\alpha_{1,2}+1)/3$.

The period 3 symmetry seems to have little implications for the
numerical integrations of the model, except perhaps in accurately
determining the structure function.

The stability of the fixed point for energy cascade in the 3D case,
$\epsilon= 1/2$, is characterized by few unstable directions, where the
corresponding eigenmodes mainly projects onto the high shell numbers,
and a large number of marginally stable directions which mainly
projects onto the inertial range. This also holds in the case with
forcing and dissipation \cite{OY}.  In the case where dissipation and
forcing are applied for some values of $\epsilon$ the Kolmogorov fixed
point can become stable. Biferale et al. \cite{biferale} show that
there is a transition in the GOY model, for $\nu=10^{-6}$ and $f=5
\times 10^{-3} \times (1+i)$, as a function of $\epsilon$ from stable
fixed point $(\epsilon < 0.38..)$, through Hopf bifurcations and via a
Ruelle-Takens scenario to chaotic dynamics $(\epsilon > 0.39..)$.

\section{Forward and backward cascades}

Until this point we have not specified which of the two conserved
quantities will cascade. Assume, in the chaotic regime where
the Kolmogorov fixed points are unstable, that there is, on average, an
input of the same size of the two quantities, $E^1$ and $E^2$, at the
forcing scale, this can of course always be done by a simple rescaling
of one of the quantities. If $N_d$ is a shell number at the beginning
of the viscous subrange, we have that $u_{N_d}/k_{N_d}\approx \nu$,and the dissipation, $D^i$, of the conserved quantity, $E^i$, can
be estimated as
 
\begin{equation}
D^i\sim \nu k_{N_d}^{\alpha_i+2)}|u_{N_d}|^2.
\label{diss}
\end{equation}
The ratio of dissipation of $E^1$ and $E^2$ scales with
$k_{N_d}$ as $D^1/D^2\sim k_{N_d}^{\alpha_1-\alpha_2}$, so that, in 
the limit $Re \rightarrow \infty$ when $\alpha_1 < \alpha_2$, there 
will be no dissipation in the viscous sub-range of $E^1$ where $E^2$ 
is dissipated. Therefore, a forward cascade of $E^1$ is prohibited and we 
should expect a forward cascade of $E^2$. For the backward cascade 
the situation is reversed, so we should expect a backward cascade of 
$E^1$.  

The situation is completely different in the 2D like and the 3D like
cases. In the 3D like models $E^2$ is not positive definite, $E^2$
(helicity) is generated also in the viscous sub-range and for the usual
GOY model we do not see a forward cascade of helicity, see, however, ref.
\cite{pdd1}. This is in agreement with the observed $k^{-5/3}$ energy
spectrum observed in real 3D turbulence corresponding to the forward 
cascade
of energy. In the 2D case we observe the direct cascade of enstrophy,
while the inverse cascade of energy is still debated. In the rest of
this paper we will concentrate on 2D like models where we will
implicitly think of $E^1=E$, with $\alpha_1=0$, as the energy and
$E^2=Z$, with $\alpha_2=\alpha>0$, as the enstrophy. With regard to the
inverse cascade of energy one must bare in mind that in 2D turbulence
the dynamics involved is probably related to the generation of large
scale coherent structures, vortices, and vortex interactions. Vortices
are localized spatially, thus delocalized in spectral space.  This is
in agreement with the estimate that 2D is marginally delocalized in
spectral space \cite{Kraichnan2}.  In the GOY model there is no spatial
structure and the interactions are local in spectral space. The model
is therefore probably not capable of showing a realistic inverse energy
cascade. We will thus only consider the forward cascade in this paper.
Figure 3 shows the scaling in the inertial sub-range of the model with
$\epsilon=5/4$ corresponding to $\alpha=2$. The cascades of the
enstrophy and energy are shown in figure 4.  It is seen that enstrophy
is forward cascaded while energy is not.

\section{Statistical description of the model}
In a statistical equilibrium of an ergodic dynamical system we will have
a probability distribution among the
(finite) degrees of freedom, assuming an ultraviolet cutoff,
of the form, $P_i\sim \exp(-BE_i^1-AE_i^2)$, where $E^1$ and
$E^2$ are the conserved quantitied, energy and enstrophy.
Thus, the temporal mean of any quantity, which is
a function of the shell velocities is given as
\begin{equation}
\overline{g}=\int \prod_i du_i g(u_1,...,u_N) \exp(-BE_i^1-AE_i^2)/
\int \prod_i du_i \exp(-BE_i^1-AE_i^2).
\end{equation}
$A$ and $B$ are Lagrange multipliers, reflecting
the conservation of energy and enstrophy when maximizing the entropy of
the system, corresponding to inverse temperatures, denoted as inverse
"energy-" and "enstrophy-temperatures" \cite{Kraichnan}. The shell
velocities themselves will in this description be independent and gaussian
distributed variables with standard deviation
$\sigma(u_i)=1/(2(Bk_i^{\alpha_1}+Ak_i^{\alpha_2}))$.  The average
values of the energy and enstrophy becomes,

\begin{eqnarray}
\overline{E_i^1}=k_i^{\alpha_1}\overline{|u_i|^2}=(B+Ak_i^{\alpha_2-
\alpha_1})^{-1}\nonumber \\
\overline{E_i^2}=k_i^{\alpha_2}\overline{|u_i|^2}=(Bk_i^{\alpha_1-
\alpha_2}+A)^{-1}.
\label{stat}
\end{eqnarray}

For $k\rightarrow 0$ we will have equipartitioning of energy,
$k_i^{\alpha_1}\overline{|u_i|^2}=B^{-1}$ and the scaling $|u_i|\sim
k_i^{-\alpha_1/2}$ and for the other branch, $k\rightarrow \infty$, we
will have equipartitioning of enstrophy
$k_i^{\alpha_2}\overline{|u_i|^2}=A^{-1}$ and the scaling $|u_i|\sim
k_i^{-\alpha_2/2}$. In the case of no forcing and no viscosity the
equilibrium will depend on the ratio $A/B$ between the initial
temperatures $A^{-1},B^{-1}$. To illustrate this we ran the model
without forcing and viscosity but with 2 different initial spectral
slopes of the velocity fields, the larger the slope the higher the
ratio of the energy temperature to the enstrophy temperature.  Figure 5
shows the equilibrium spectra for $\epsilon=5/4, \nu=f=0$, in the cases
of initial slopes -1, -0.8. The full lines are the equilibrium
distribution given by (\ref{stat}) for $A/B=10^2$ and $A/B=10^{-2}$
respectively.

\section{Distinguishing cascade from statistical equilibrium}
For the forward enstrophy cascade the spectral slope is $-(\alpha
+1)/3$ and the enstrophy equipartitioning branch has spectral slope
$-\alpha/2$. Thus for the 2D case where $\alpha=2$ we cannot distinguish
between statistical (quasi-) equilibrium and cascading.  This was
pointed out by Aurell et al. \cite{aurell} and it was argued that the
model can be described as being in statistical quasi-equilibrium with
the enstrophy transfer described as a simple diffusion rather than an
enstrophy cascade. This coinciding scaling is a caviate of the GOY
model not present in the real 2D flow where the statistical
equilibrium energy spectrum scales as $k^{-1}$ and the cascade energy
spectrum scales as $k^{-3}$. For other values of $\alpha$ the scaling
of the two cases are different, see figure 6.  This figure represents
the main message of this paper. First axis is the parameter $\epsilon$,
along the line shown in fig. 1, defining the spectral ratio between the
two inviscid invariants. Second axis is the scaling exponent $\gamma$.
The horizontal dashed line $\gamma=1/3$ is the Kolmogorov scaling
exponent for energy cascade. The full curve is the scaling exponent for
the enstrophy cascade, and the dotted curve corresponds to the
enstrophy equipartitioning.

All the 3D like models (asterisks in figure 6) are near energy 
cascade scaling
(dashed line). Statistical equilibrium corresponds
to the line $\gamma=0$.  The bold line piece, $0<\epsilon < 0.39...$,
represents parameter values where the Kolmogorov fixed point is
stable \cite{biferale}. The scaling for $\epsilon > 0.39...$  is
slightly steeper than the Kolmogorov scaling, which is attributed to
intermittency corrections originating from the viscous
dissipation \cite{p+v+j}. It seems as if there is a slight trend showing
increasing spectral slopes for increasing $\epsilon$.

For the 2D like models the scaling slope is also everywhere on or
slightly above both the cascade - and the equilibrium slopes (diamonds
in the figure). The classical argument for a cascade is that given an
initial state with enstrophy concentrated at the low wave-number end of
the spectrum, the enstrophy will flow into the high wave-numbers in
order to establish statistical equilibrium.  The ultra-violet
catastrophe is then prevented by the dissipation in the viscous
sub-range.  Therefore, we cannot have a non-equilibrium distribution
with more enstrophy in the high wave-number part of the spectrum than
prescribed by statistical equilibrium since enstrophy in that case
would flow from high - to low wave-numbers. This means that the
spectral slope in the inertial sub-range always is above the slope
corresponding to equilibrium (dotted line in figure 6). Consequently,
the 2D model with $\epsilon=5/4$ separates two regimes,
$1<\epsilon<5/4$ where enstrophy equilibrium is achieved and
$5/4<\epsilon<2$ where the enstrophy is cascaded through the inertial
range.

In figure 7 the spectra and the cascades are shown for different values
of $\epsilon$. The model was run with 50 shells and forcing on shell
number 15 for $2 \times 10^4$ time units and averaged. Even then there
are large fluctuations in the cascades not reflected in the spectra. 
The large differences in the absolut values for the cascades, $Pi$,
is a reflection of the scaling relation (\ref{diss}).

We
interpret the peaks around the forcing scale for $\epsilon=11/10$ as
statistical fluctuation and the model shows no cascade. For
$\epsilon>5/4$ we see an enstrophy cascade and what seems to be an
inverse energy cascade. However, we must stress that we do not see a
second scaling regime for small $n$ corresponding the inverse
cascade.  Note that for $\epsilon=2$ energy and enstrophy are identical
and we have only one inviscid invariant.  So if a regime of inverse
energy cascading existed in parameter space near $\epsilon=2$ 
the scaling exponents will be almost identical and coincide at
$\epsilon=2$.

The two regimes corresponding to equipartitioning and cascade can be
understood in terms of timescales for the dynamics of the shell
velocities.  A rough estimate of the timescales for a given shell $n$,
is from (\ref{dyn}) given as $T_n\sim (k_nu_n)^{-1}\sim
k_n^{\gamma-1}$. Again $\epsilon=5/4$, corresponding to $\gamma=1$,
becomes marginal where the timescale is independent of shell number.
For $\epsilon< 5/4$ the timescale grows with $n$ and the fast
timescales for small $n$ can equilibrate enstrophy among the degrees of
freedom of the system before the dissipation, at the "slow" shells, has
time to be active. Therefore these models exhibits statistical
equilibrium.  For $\epsilon>5/4$ the situation is reversed and the
models exhibits enstrophy cascades. Time evolutions of the shell
velocities are shown in figure 8, where the left columns show the
evolution of a shell in the beginning of the inertial subrange and the
right columns show the evolution of a shell at the end of the inertial
subrange.  This timescale scaling might also explain why no inverse
cascade branch has been seen in the GOY model.  The timescales at the
small wave-number end of the spectrum, with the dissipation or drag
range for inverse cascade, is long in comparison with the timescales of
the inertial range of inverse cascade. Therefore a statistical
equilibrium will have time to form.  The analysis suggests that
parameter choices $\epsilon > 5/4$ might be more realistic than
$\epsilon=5/4$ for mimicing enstrophy cascade in real 2D turbulence.

\section{Intermittency corrections}
The numerical result that the inertial range scaling has a slope
slightly higher than the K 41 prediction, is not fully understood.
This is attributed to intermittency corrections originating from
the dissipation of enstrophy in the viscous subrange.

The evolution of the shell velocities in the viscous sub-range is
intermittent for $\epsilon>5/4$, where the PDF's are non-gaussian,
while the PDF's for $\epsilon=5/4$ are
gaussian in both ends of the inertial sub-range, see figure 9.  
The deviation from the Kolmogorov scaling is expressed through the
structure function, $\zeta (q)$ \cite{p+v+j}. The structure 
function is defined through the scaling of the moments of the 
shell velocities;

\begin{equation}
\overline{|u_n|^q}\sim k_n^{\zeta(q)}=k_n^{-q\gamma -\delta\zeta (q)}
\end{equation}
where $\delta\zeta (q)$ is the deviation from Kolmogorov scaling.  The
structure function, $\zeta (q)$, and $\delta\zeta (q)$ for
$\epsilon=11/10,5/4,3/2,7/4,2$ are shown in figure 10. For
$\epsilon>5/4$ there are intermittency corrections to the scaling in
agreement with what the PDF's show.

We know of no analytic way to predict the intermittency corrections
from the dynamical equation. Our numerical calculations suggest that
the intermittency corrections are connected with the differences in
typical timescales from the beginning of the inertial sub-range, where
the model is forced, to the viscous sub-range.  The ratio of timescales
between the dissipation scale and the forcing scale can be estimated
by; $T_{\nu}/T_f\approx \lambda^{ \Delta N (1+\gamma )}$, where $\Delta
N$ is the number of shells beween the two.  Figure 11 (a) shows the
numerical values of $\delta\zeta (10)$ as a function of $\epsilon$ and
figure 11 (b) shows $log_2(T_{\nu}/T_f)$ as a function of $\epsilon$.
The vertical line indicates the crossover between statistical
equilibrium and cascading.
We must stress that caution should be taken upon drawing conclusions
from this since the authors have no physical explanation of the
appearent relationship.

\section{Summary}

The GOY shell model has two inviscid invariants, which govern the
behavior of the model. In the 2D like case these corresponds to the
energy and the enstrophy of 2D turbulent flow. In the model we can
change the interaction coefficient, $\epsilon$, and tune the spectral
ratio of enstrophy to energy, $Z_n/E_n=k_n^\alpha$. For $\alpha>2$ we
can describe the dynamics as being in statistical equilibrium with two
scaling regimes corresponding to equipartitioning of energy and
enstrophy respectively. The reason for the equipartitioning of
enstrophy in the inertial range (of forward cascading of enstrophy) is
that the typical timescales, corresponding to eddy turnover times, are
growing with shell number, thus the timescale of viscous dissipation is
large in comparison with the timescales of non-linear transfer. Thus,
this choice of interaction coefficient is completely unrealistic for
mimicing cascades in 2D turbulence. For $\alpha<2$ the model shows
forward cascading of enstrophy, but we have not identified a backward
cascade of energy. The usual choice $\epsilon=5/4$, $\alpha=2$ is a
borderline and we suggest that $\alpha<2$ in respect to mimicing
enstrophy cascade might be more realistic. We observe that the dynamics
becomes more intermittent when $\alpha<2$, in the sense that the
structure function deviates more and more from the Kolmogorov
prediction. For $\epsilon=2$ we have $\alpha=0$, thus energy and
enstrophy degenerates into only one inviscid invariant, this point
could then be interpreted as a model of 3D turbulence. However, as is
seen from (\ref{pi1}), in this case the fluxless fixed point is the one
surviving, but as is seen in figure 7, bottom panels, this model also
shows cascading.  This choice for 3D turbulence model could shed some
light on the dispute of the second inviscid invariant (helicity) being
important \cite{benzi} or not \cite{procaccia} for the deviations from
Kolmogorov theory, work is in progress on this point.

\section{Acknowledgements}We would like to thank Prof. A. Wiin-Nielsen
for illuminating discussions. This work was supported by the 
Carlsberg Foundation.

\newpage

\begin{figure}[htb]
\epsfxsize=10cm
\epsffile{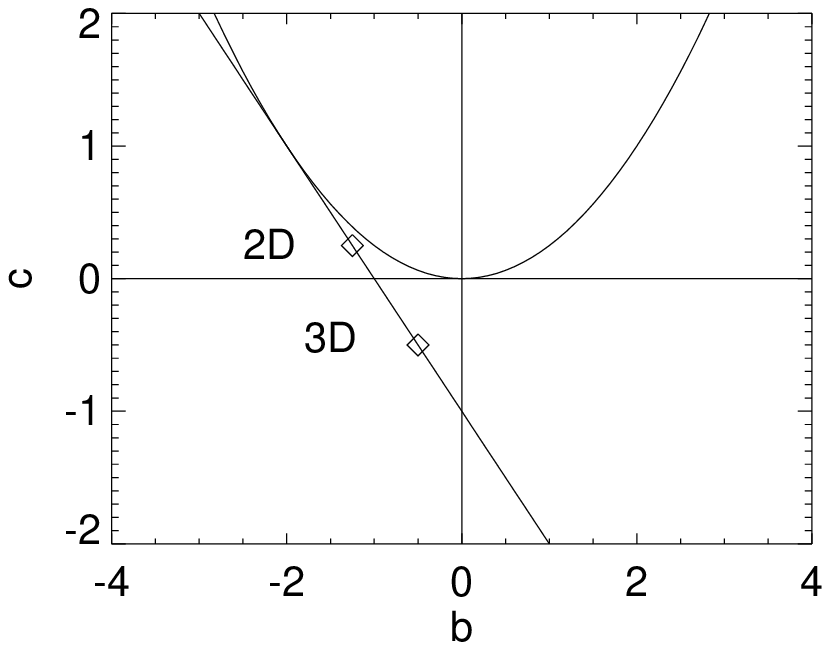}
\caption[]{ 
\label{1}The (b, c) parameter space. The line indicates where the
one conserved integral is the energy, defined as $E=1/2 \sum |u_n|^2$
}
\end{figure}

\begin{figure}[htb]
\epsfxsize=10cm
\epsffile{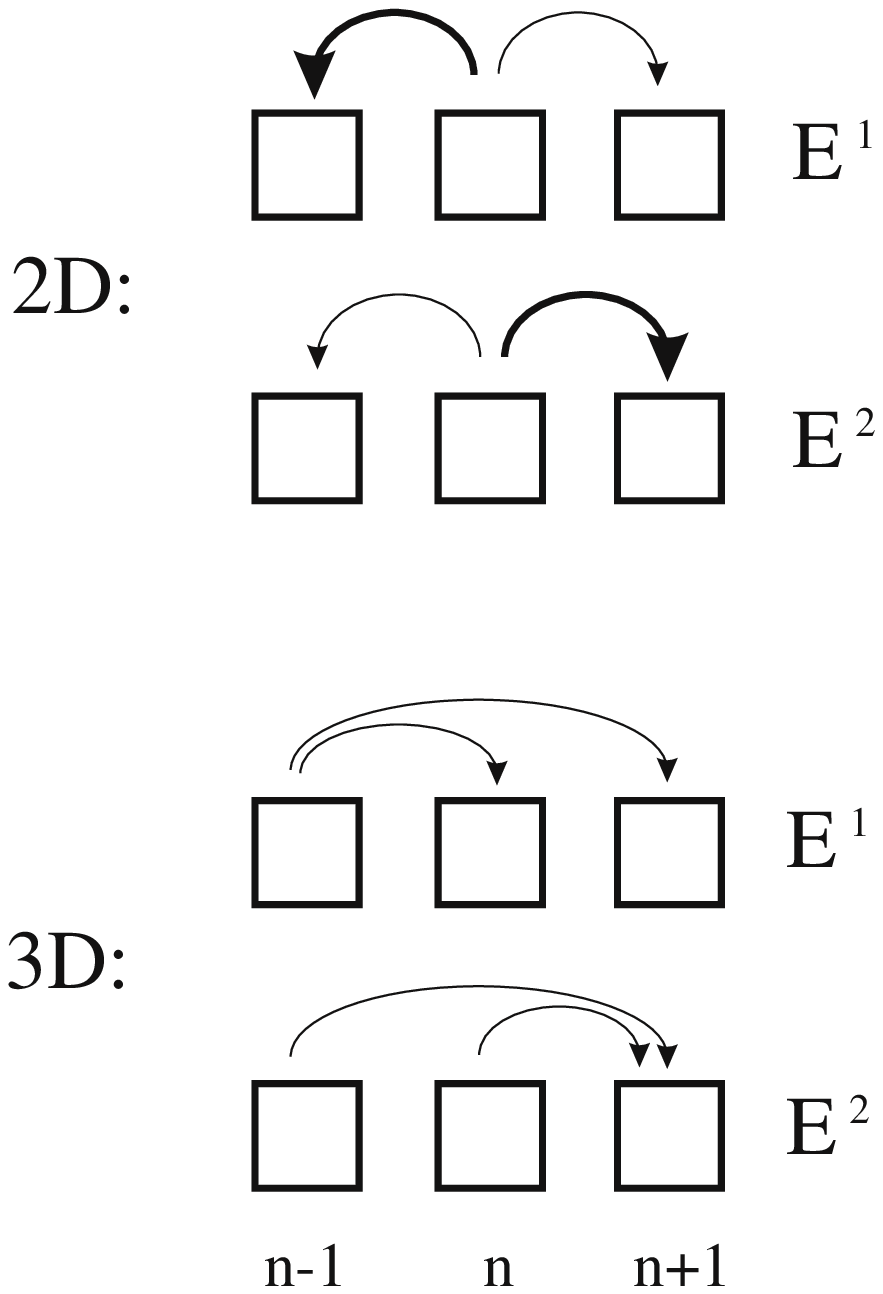}
\caption[]{ 
\label{2a}The shell triad interactions. Arrows indicates transfer
of energy, $E^1$, and enstrophy, $E^2$, for the 2D case,
$\epsilon=5/4$, and energy, $E^1$, and helicity, $E^2$, for 3D,
$\epsilon=1/2$,  case. The thickness of the arrows indicates the
strength of the transfer.
}
\end{figure}

\begin{figure}[htb]
\epsfxsize=10cm
\epsffile{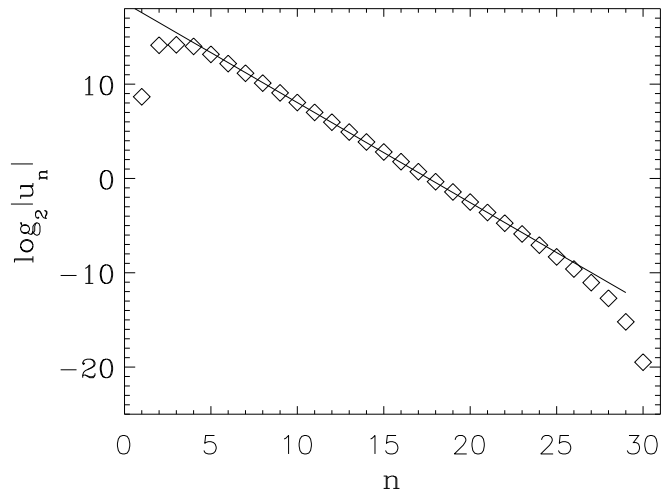}
\caption[]{ 
\label{3}The mean value of the shell velocities as a function of
shell number on a logarithmic scale (base $\lambda$), for the 2D
case, $\epsilon=5/4, k_0= \lambda^{-4}, \lambda=2,n=30, \nu= 10^{-
16}, f_n= 5 \times 10^{-3} \times (1+i) \delta_{n,4}$. The
model was run for $4.2 \times 10^{4}$ time-units.
}
\end{figure}

\begin{figure}[htb]
\epsfxsize=10cm
\epsffile{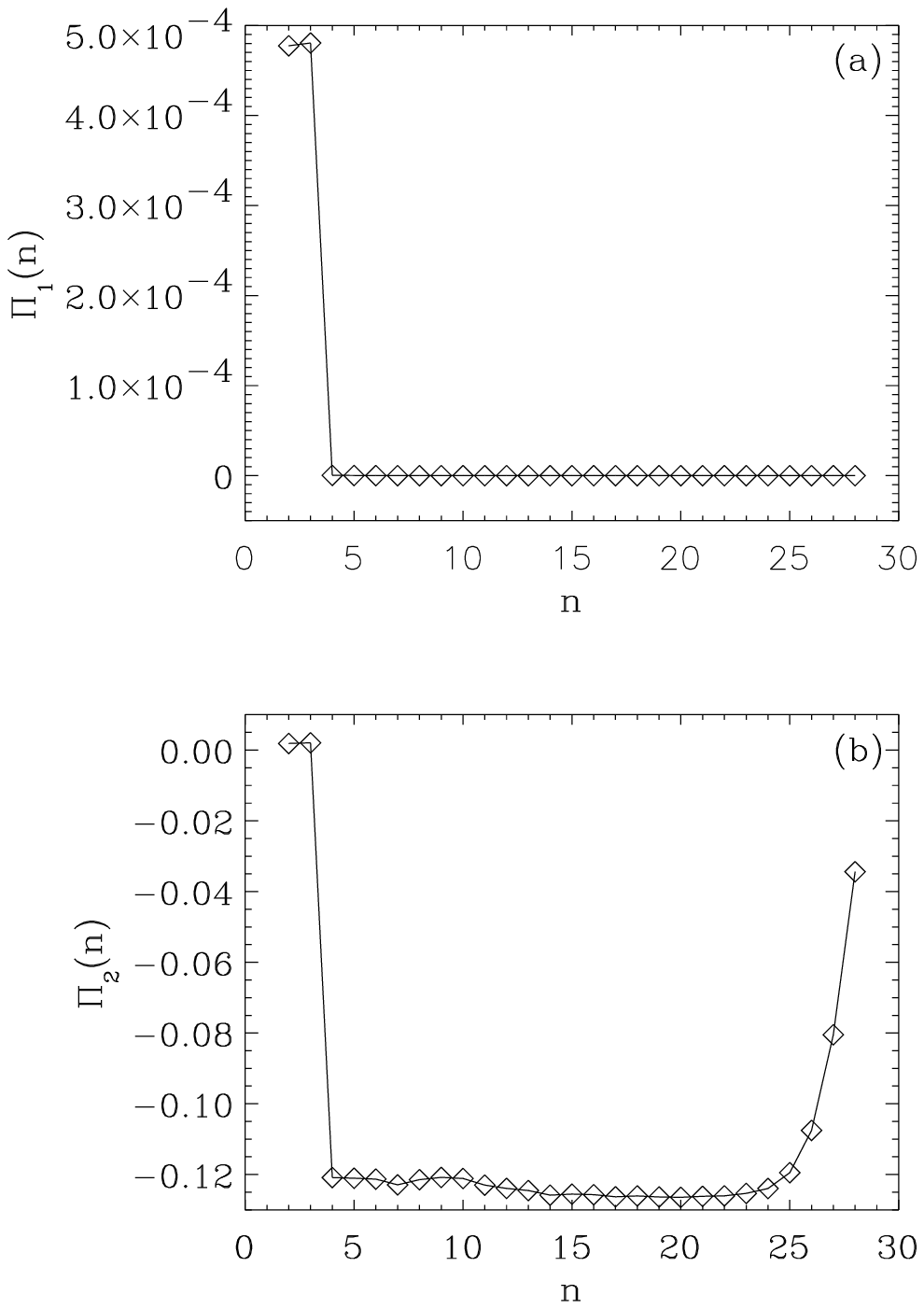}
\caption[]{ 
\label{4}The mean values of, (a),  the energy flux, $\Pi_1$, 
and (b) the enstrophy flux, $\Pi_2$. 
}
\end{figure}

\begin{figure}[htb]
\epsfxsize=10cm
\epsffile{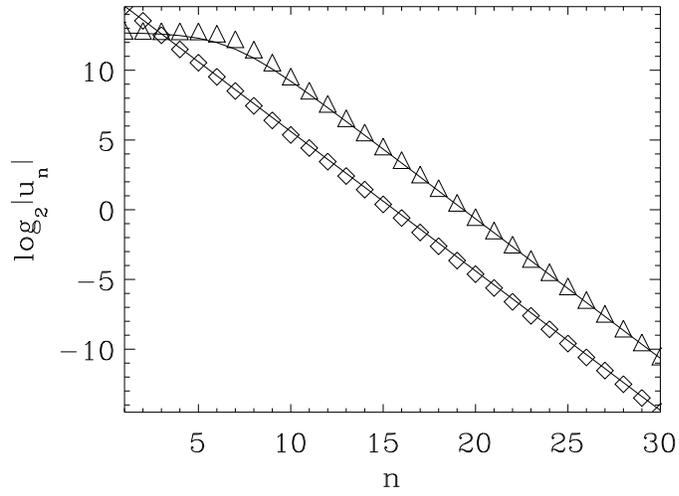}
\caption[]{ 
\label{5}Same as figure 3, but with $n=30, \nu=f=0$. Diamonds
corresponds to an initial spectral slope of -1.0, that is a high value
of A/B.  The corresponding curve is statistical equilibrium
distribution for $A/B=10^2$.  Triangles corresponds to an initial
spectral slope of -0.8, that is a lower value of A/B.  The curve is
statistical equilibrium distribution for $A/B=10^{-2}$.
}
\end{figure}

\begin{figure}[htb]
\epsfxsize=12cm
\epsffile{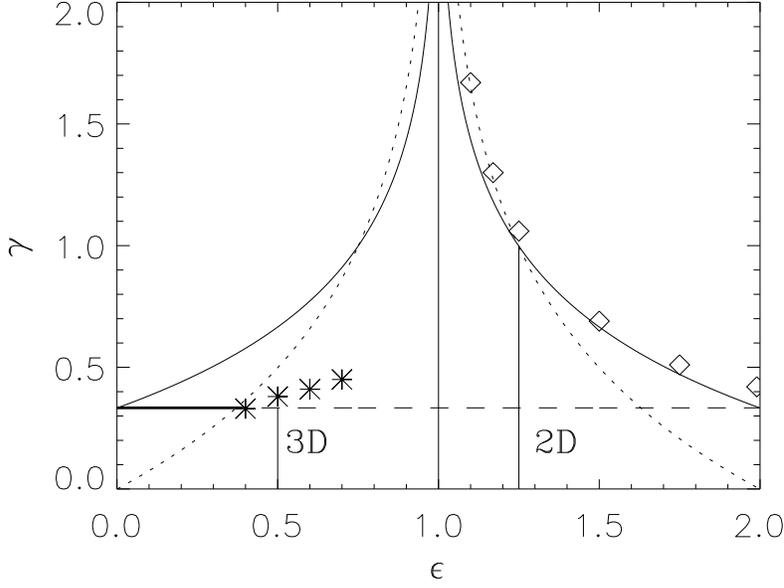}
\caption[]{ 
\label{6}The spectral slope $\gamma$ as function of $\epsilon$.
The horizontal dashed line is the Kolmogorov scaling for energy
cascade. The full curve is the scaling exponent for the enstrophy (or
in 3D like case helicity) cascade, and the dotted curve corresponds to
the enstrophy (helicity) equipartitioning.  All the 3D like models
shows energy cascade (equipartitioning corresponds to the line $\gamma=0$).
The bold line piece, $0<\epsilon < 0.39...$, represents parameter
values where the Kolmogorov fixed point is stable \cite{biferale}.
The diamonds are model run with $n=50, k_0= \lambda^{-4}, \lambda=2,
f_n= 5 \times 10^{-4} \times (1+i) \delta_{n,15}, 
(\epsilon =11/10,\nu = 5 \times 10^{-27}, \nu'=100),
(\epsilon =117/100,\nu = 5 \times 10^{-27}, \nu'=100),
(\epsilon =5/4,\nu = 5 \times 10^{-25}, \nu'=100),
(\epsilon =3/2,\nu = 5 \times 10^{-23}, \nu'=100),
(\epsilon =7/4,\nu = 5 \times 10^{-23}, \nu'=100),
(\epsilon =2,\nu = 5 \times 10^{-20}, \nu'=100).$
The stars are model runs with $n=19, k_0= \lambda^{-4}, \lambda=2,
 f_n= 10^{-4} \times (1+i) \delta_{n,4},\nu = 10^{-6}, \nu'=0, \epsilon
=1/2,6/10,7/10$.  The 3D like models shows Kolmogorov scaling, with
deviations due to intermittency corrections, and energy cascading.  The
2D like models shows a cross-over at $\epsilon=5/4$ between
statistical equilibrium, $1<\epsilon<5/4$, and enstrophy cascading,
$5/4<\epsilon<2$.
}
\end{figure}

\begin{figure}[htb]
\epsfxsize=10cm
\epsffile{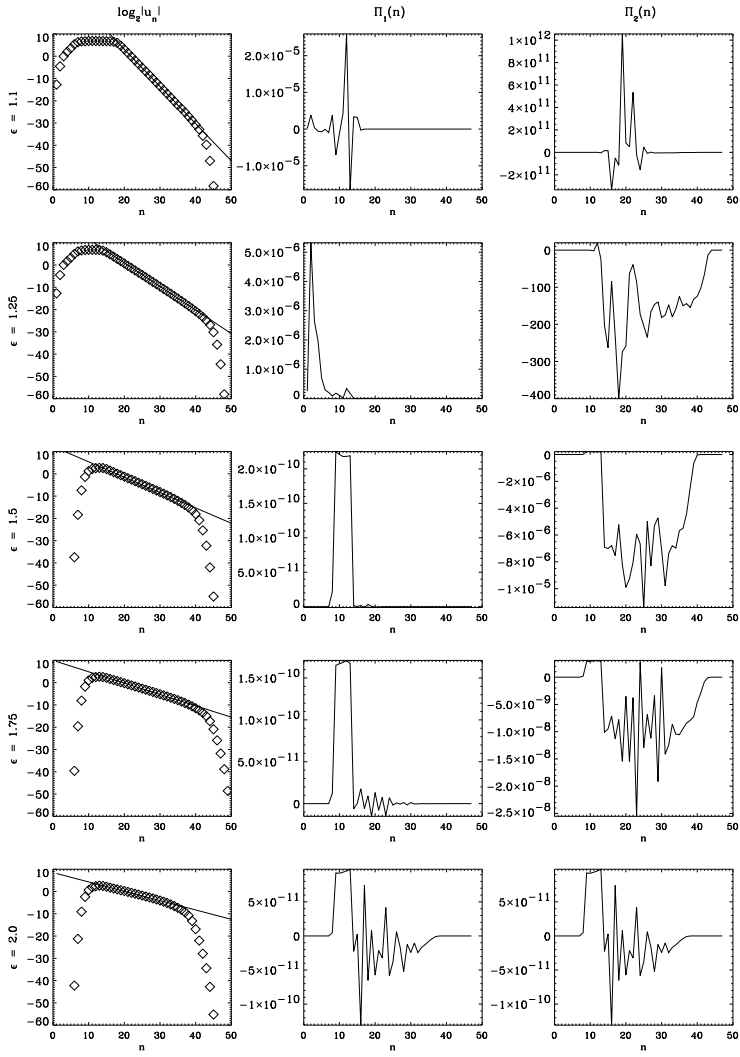}
\caption[]{ 
\label{7}Same as figures 3 and 4 for $\epsilon=11/10, 5/4, 3/2, 
7/4, 2$. The spectral slopes are shown in figure 6 (diamonds).
}
\end{figure}

\begin{figure}[htb]
\epsfxsize=10cm
\epsffile{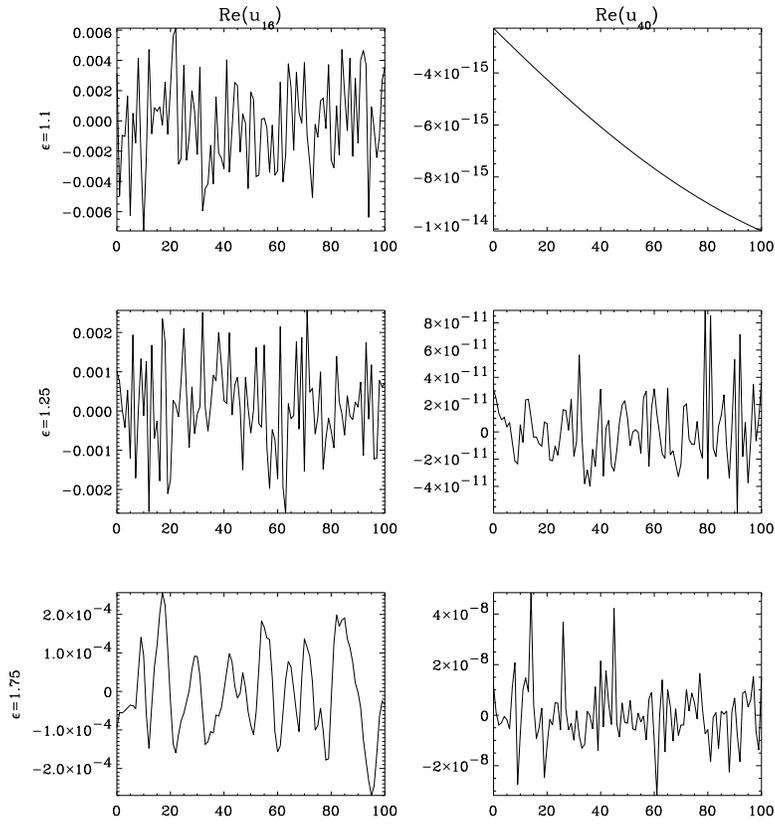}
\caption[]{ 
\label{8}Time evolution of shell velocities in the beginning
and the end of the inertial subrange. The typical timescale of
shell $n$ scales as $T_n\sim (k_n |u_n|)^{-1} \sim k_n^{\gamma-1}$.
Note that for $\epsilon=5/4$ the timescale is the same for all shells.
}
\end{figure}

\begin{figure}[htb]
\epsfxsize=10cm
\epsffile{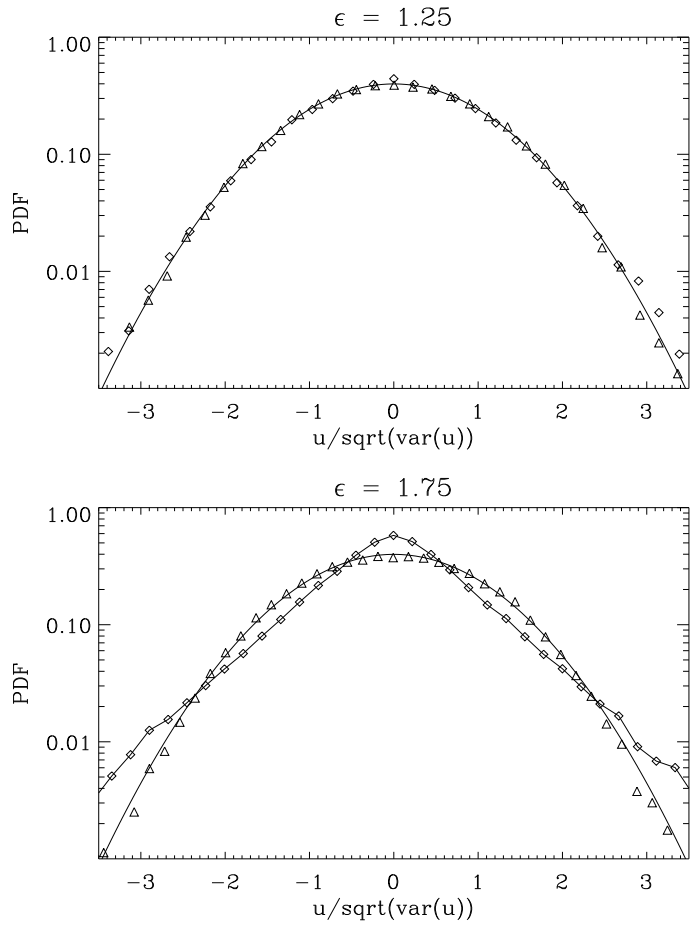}
\caption[]{ 
\label{9}Probability density functions for real and imaginary
parts of shell velocities. The curves are gaussians with the same 
variance.
}
\end{figure}

\begin{figure}[htb]
\epsfxsize=10cm
\epsffile{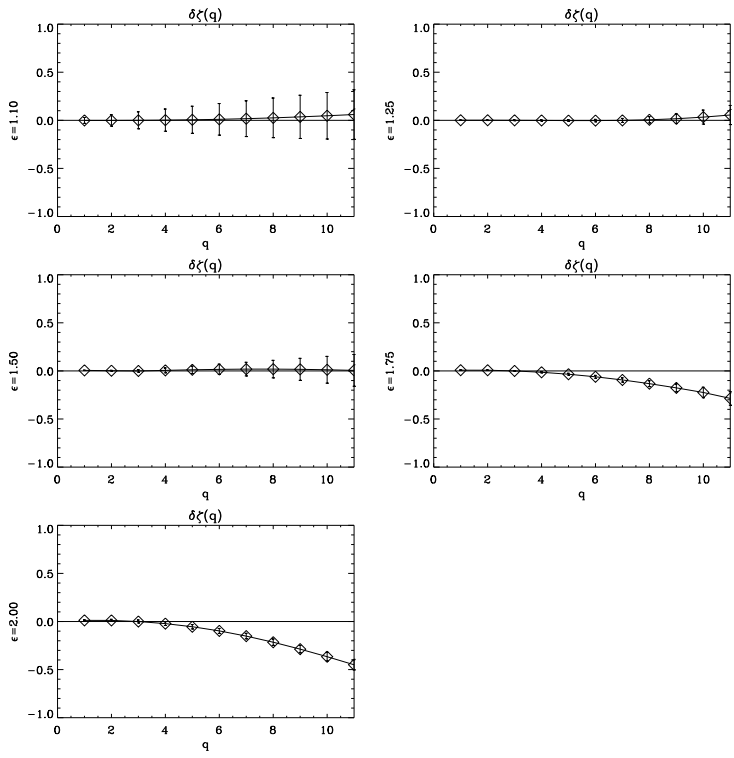}
\caption[]{ 
\label{10}The deviation of the structure function from Kolmogorov
scaling for 
$\epsilon=11/10, 5/4, 3/2, 7/4, 2$.
}
\end{figure}

\begin{figure}[htb]
\epsfxsize=10cm
\epsffile{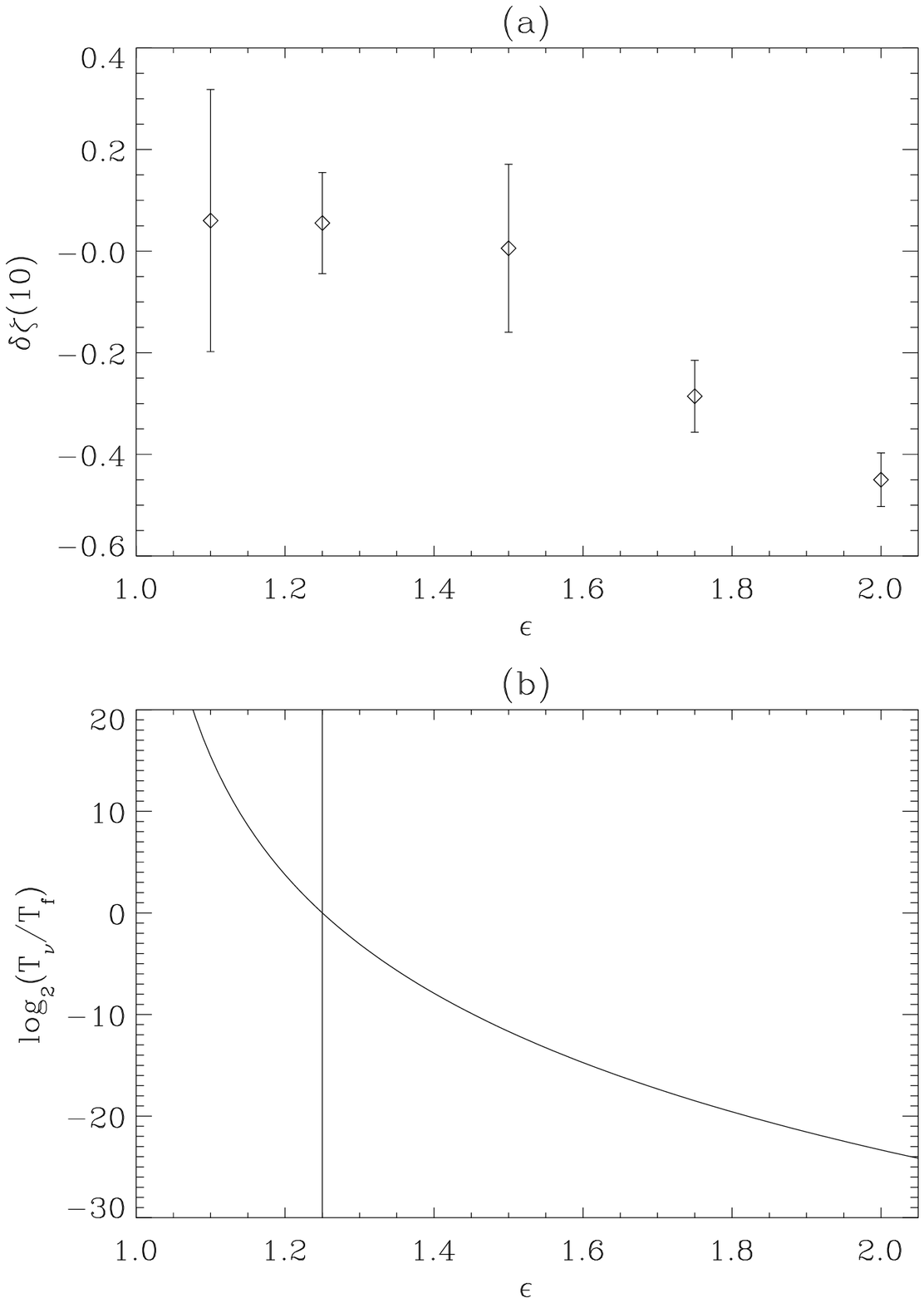}
\caption[]{ 
\label{11} (a) Numerical values of $\delta\zeta (10)$ as a
function of $\epsilon$, the error bars represents one standard
deviation.  (b) Ratio of typical timescales between dissipation scale
and forcing scale, $log_2(T_{\nu}/T_f)$ as a function of $\epsilon$.
The vertical line indicates the crossover between statistical
equilibrium and cascading.
}
\end{figure}

\begin{thebibliography}{xx}

\bibitem{Gledzer} E. B. Gledzer, Sov. Phys. Dokl, 18, 216 (1973).
\bibitem{GOY} M. Yamada and K. Okhitani, J. Phys. Soc. of Japan 56,
4210 (1987); Progr. Theo. Phys. 79, 1265 (1988).
\bibitem{benzi}L. Kadanoff, D. Lohse, J. Wang and R. Benzi, Phys. Fluids 7, (1995).
\bibitem{aurell}E. Aurell, G. Boffetta, A. Crisanti, P. Frick, G. Paladin and A. Vulpiani, Phys. Rev. E, 50, 4705, (1994).
\bibitem{euro}P. Frick and E. Aurell, Europhys. Lett. 24, 725, 
(1993).
\bibitem{biferale}L. Biferale, A. Lambert, R. Lima, G. Paladin, Physica D, 80, 105 (1995).
\bibitem{pdd1}P. D. Ditlevsen, to be published
\bibitem{OY} M. Yamada and K. Okhitani,Phys. Rev. Lett. 60,983 (1988)
\bibitem{schorghofer}N. Sch\"orghofer, L. Kadanoff and D. Lohse, 
{\sl preprint 1995}.
\bibitem{Kraichnan}R. H. Kraichnan and D. Montgomery, Rep. Prog. 
Phys., 43, 547, (1980).
\bibitem{Kraichnan2}R. H. Kraichnan, J. Fluid Mech., 47, 525, (1971).
\bibitem{p+v+j}M. H. Jensen, G. Paladin and A. Vulpiani, Phys. Rev. A, 43, 798, (1991).
\bibitem{procaccia}O. Gat, I. Procaccia and R. Zeitak, {sl preprint, 1994}
\end{thebibliography}
\end{document}